\documentclass{article}
\usepackage[utf8]{inputenc}
\usepackage{xcolor}
\usepackage{geometry} 
\usepackage{xspace} 
\usepackage{ulem} 
\usepackage{hyperref}
\usepackage{authblk}

\title{MolSSI and BioExcel Workflow Workshop 2018 Report}
\author[1]{Levi N. Naden}
\author[1]{Sam Ellis} 
\author[1]{Shantenu Jha}

\affil[1]{The Molecular Sciences Software Institute}
\date{April 2019}

\newcommand{\work}{Workflow System\xspace}
\newcommand{\works}{Workflow Systems\xspace}

\begin{document}
	
\maketitle	

\abstract{Workflows in biomolecular science are very important as they are
intricately intertwined with the scientific outcomes, as well as algorithmic
and methodological innovations. The use and effectiveness of workflow tools to
meet the needs of the biomolecular science community is varied. MolSSI
co-organized a biomolecular workflows workshop in December 2018 with the goal
of identifying specific software gaps and opportunities for improved workflow
practices. This report captures presentations and discussion from that
workshop. The workshop participants were primary tools developers, along with
"neutral observers" and some biomolecular domain scientists. After
contextualizing and motivating the workshop, the report covers the existing
roles and emerging trends in how workflow systems are utilized. A few
recurring observations are presented as recommendations for improving the use
and effectiveness of workflow tools. The tools presented are discussed in
Appendix~\ref{app:tool_descriptions}.}


\section{Workshop Motivation, Context and Overview}\label{sec:introduction}

Most scientific endeavors today require computational campaigns with multiple
tasks and distinct runs, e.g., parameter optimization, exploratory runs and
sensitivity analysis; as opposed to just a single task or a predefined number
of execution or invocations of those tasks.  Arguably, nowhere is the
importance of workflows greater than in biomolecular sciences where the
scientific outcomes are intricately intertwined  with the ability to execute
workflows and campaigns successfully. Furthermore, the algorithmic and
methodological innovations are tied to the successful formulation and
execution of workflows \cite{lindorff}.

In response to requirements of automation, scale and sophistication, there
have been a variety of projects aimed at developing workflow systems -- some
customized, some general purpose. There have been significant advances in the
state-of-the-theory and practice in workflows, however, the state of workflow
development, execution and extension leaves much scope for improvement.

To take just one example, there are in excess of 230 purported workflow
systems \cite{cwl_workflow_list} enumerated by one source. From a technical
point of view, an important question is why are there so many?  It is
difficult to discern all underlying causes for so many workflow systems, the
proliferation highlights problems of (i) missing building blocks and
abstractions leading to redundancy in implementation and insufficient reuse,
(ii) inadequate attention to sustainability, (iii) possibly inadequate
ontological framework and terminology to organize and distinguish workflow
tools from each other (How might the unique, necessary or important ones be
sustained?). From the users point of view, which of the 230+ systems might be
useful or the best starting point? What factors should be considered when
assessing which workflow system to adopt?


An aim of this workshop was to ensure the Biomolecular Simulations (BMS)
community is both aware and able to leverage these advances where possible,
while avoiding the pitfalls that "general" workflow system developers have
encountered (e.g., proliferation and partial solutions) and ultimately
enhancing the BMS community's repertoire and use of workflow tools. To this
end, the workshop organizers decided to bring together workflow tools and
system development projects from both sides of the Atlantic with the goal to
have community agreement on what solutions / systems  are most needed by the
broader BMS researchers and thus have the best chance of being long-term
sustainable products. Another explicit objective of the workshop was to help
determine where MolSSI -- an NSF funded software institute, could contribute
to maximize impact while avoiding both redundancy and competing with its
stakeholders.

The needs of the BMS community are vast and in order to make
tangible progress it was decided that the workshop would focus, initially, on
two of the most representative workflow patterns (in terms of frequency of
occurrence), as well as from which more “complex” workflows could seemingly be
constructed. These decisions were made after seeking and receiving more than
half a dozen use cases from stakeholders.

The organizers had the clear intention and mission that the workshop would not
serve as  a place for projects to make “sales pitches”. Also, it was the
intent to not host yet another "talking shop". Thus the organizers decided not
to invite users to this first workshop. Rather, the workshop organizers wanted
to bring participants willing and able to discuss both the strengths and
weaknesses of their approaches, what opportunities could come for synergising
with other work, and where there are still gaps in provision and unmet needs.
Thus, the focus on workflow systems and tools developers.

Each presenting team was asked to prepare a 45 minute presentation to address
the  following broad outline:

\begin{itemize}
\item The Science Drivers: what types of scientific problems your workflow solution has been designed to help overcome?
\item An introduction to your project/tool – a bit of background, history, state of development, broad-brush description of how it works and what it does.
\item Examples/major successes in application to technology/science problems
\item Major problems/challenges in application to technology/science problems
\item A wish-list of what you would like to get out of this workshop.
\end{itemize}

Following general overview presentations  on Day 1, there were extensive
opportunities on subsequent days to dive deeper into each tool's / system's
design, code, challenges and approaches.  The workshop was deliberately under
structured to allow participants the opportunity and freedom to self-organize
and self-select the best approach to reach the desired goals. Links to the
presentation and written responses from the presenters can be found in the 
references \cite{ref:workshop_materials}.
 
We close the introduction with terminology to ensure proper context and
consistency. "Workflows" are  a way to express dependencies of units that
comprise the multi-component and multi-stage application. These dependencies
codify scientific semantics of the application.  Often the applications have
well defined dependencies; sometimes the dependencies must be resolved at
runtime. Further, the same workflow can be expressed in different ways, e.g.,
constructed by a directed acyclic graph (DAG) or non-DAG representations; similarly,
there are different algorithms and approaches to resolve dependencies which 
are typically encoded within the workflow engine. Workflows are distinct from 
workloads: "workloads" are a set of units that can be executed concurrently, 
wherein the units are devoid of semantic context. The tool which executes 
a given workflow or workflows we define as a "\work."

We distinguish a workflow from a computational campaign. The computational
activity to achieve a defined objective (i.e., computing binding affinities
for multiple drug candidates) is referred to as a "computational campaign." In
contrast, a workflow is a specific and well-defined execution of a set of
computational tasks, many of which could be aggregated to achieve the
objectives of the computational campaign. For example, a computational
campaign could be the entire activity of determining an optimal drug design.
The campaign could be comprised of multiple instances of a workflow to compute
binding affinity of a set of drug candidates -- either individually computed
or in an ensemble mode.



\section{Enhancing Workflows with \works} \label{sec:tool_summary}

        A workflow consists of a set of computational tasks that are executed in a well-defined order. Each step consists of an executable task and may contain a set of inputs, a set of outputs, 
        and/or execution parameters.
        A step may have dependencies on other stages in the workflow and can only be executed once all of its dependencies have been completed.
        A user can manually execute these steps by performing each computational task in sequence and ensuring all data and parameters move between the tasks smoothly.
        An alternative is to construct a workflow within a \work to handle the execution of the steps automatically.
        There are a number of reasons to consider utilizing a \work to perform scientific research.
        \begin{itemize}
    	\item \textbf{Experimental Complexity}: Constraints on resources, such as time, money, policy restrictions, or reliability eliminate the possibility to perform the experiment manually. \works can remove human interaction during an experiment, reducing the drain on resources and increasing the reliability of the results.
    
    	\item \textbf{Data Control Flow}: Some workflows contain branching decisions during their operation. Manually executing such a workflow requires user interaction, which can delay the experiment, costing valuable resources. Many \works can be programmed to automatically handle branching workflows. 
    	
        \item \textbf{Reproducibility}: A key component in reviewing the quality of scientific research is the ability to reproduce the results of an experiment. A \work packages the process of the experiment, allowing it to be performed multiple times without the need to recreate it exactly.
    
    	\item \textbf{Provenance}: \works contain all of the steps used to perform the scientific experiment, ensuring that it is simple to determine which steps were taken to produce a given set of results.
    
    	\item \textbf{Dissemination}: A structured workflow makes a scientific process more accessible to other users. \works provide a more standard format for your scientific process that contains all the information needed to reproduce your experimental results. Accessibility and ease of use are large contributing factors to the adoption of processes.
    
    	\item \textbf{Resources}: Many \works improve the usage of computing resources, such as memory, disk space, or computation nodes.
    	
        \item \textbf{Deployment}: \works can aid in the deployment of a workflow to additional computing resources, such as cloud based computers or super computing clusters.
    
    \end{itemize}
    
    	Once a user has determined to adopt a \work to perform their work, it is necessary to determine which one will best suit their particular needs.
    	There are a large number of \works available to use, covering applications ranging from very specific use cases to broad, general workflows.
    	
    	Directly suggesting a \work to use is a difficult task due to the intricacies involved in each scientific problem.
    	However, there are some shared factors that any user should be aware of when they are researching different \works to use to perform their research.
    	The following list provides a base set of elements, as a starting point, to consider when selecting a \work. Not all of these items may apply to every user, and a user should 
    	determine their relative importance on a case-by-case basis.
    	\begin{itemize} \label{list:workflow_considerations}
    		\item \textbf{Barrier of Entry}: 
    		\begin{itemize}
    			\item \textbf{Ease of Use}: How difficult is it to set up and apply the \work to your scientific need? What is the quality of the documentation provided by the \work? How can you interface other applications with the \work?
    			\item \textbf{Support Tools}: How active and responsive is the development team for the \work? What computational platforms does the \work support?
    			\item \textbf{User Interface}: Which programming languages are supported by the \work? Does the \work provide a Graphical User Interface? How is the workflow represented within the provided tools?
    		\end{itemize}
    	
    		\item \textbf{Reliability}: How fault tolerant is the \work? What features does it contain to notify you of errors and allow you to debug the problem? Are you restricted to a pre-built set of error checking capabilities or are you allowed to customize them?
    		
    		\item \textbf{Reproducibility}: What information is preserved by the workflow produced through the tool? How is the workflow represented outside of the tool? What is the portability of the workflow produced by the tool and can it be imported into other systems?
    
    		\item \textbf{Execution Options}: What type of tasks are you trying to execute: binary executables, functions and method calls, or a combination of the two? Do you require specific execution libraries such as MPI or OpenMP?
    		
    		\item \textbf{Task Dimensionality}: What are the qualities of your tasks? How many jobs do you need to run? What resources do your jobs require, such as memory or processing requirements? How parallel are the tasks you are trying to run? Do your tasks operate independently or do they require communication?
    		
    		\item \textbf{Data Management}: Will you manually handle data transport between computing resources and your data storage or do you require the \work to handle the data? What are your file management requirements?
    		
    	\end{itemize}
    	
    	A selection of groups performing research in \works were involved in the discussions at the workshop, a brief summary of each tool is available in Appendix~\ref{app:tool_descriptions}.

\section{Emerging Trends Observed from the Workshop}\label{sec:discussion_breakdown}

    Several common ideas were observed after the individual \works presented their tools. The workshop collected these commonalities and felt that
    many of them were worth documenting to find easier means to link tools together, avoid regular pitfalls for current and future developers, 
    help users find the correct tool to adopt for their problems, and identify the future direction \works.
    

    \paragraph{Bigger is not better: } The most important trend is that
    individual \work developers 
    are moving from solving the entire scientific pipeline to developing tools focused on handling small, singular tasks well, such as figuring out how to distribute arbitrary jobs on HPC systems. This choice comes 
    with the conscious decision to not be experts at individual scientific problems; e.g.\ handling an entire protein-ligand binding pipeline of 
    file preparation, simulation, and analysis is forgone in favor of just one of the stages. 
    Overall, this was received as a good thing as it indicates the packages are moving towards 
    more modular tools, specializing in a single tasks, which can then be integrated together at a user's discretion to best suit the user's needs.
    An important note here is that within the scope of a single class of task, e.g. HPC distribution, input file preparation, logical flow between tasks, 
    etc., the consensus was no one tool will ever be the dominant because every scientific problem has unique requirements.
    In fact, the consensus was that many of these \works can work in concert with one another, taking advantage of the specific problems each 
    one addresses.
    Therefore, it is important when choosing a \work to be mindful of the decision metrics outlined in Section~\ref{sec:tool_summary} and 
    we refer readers to Appendix~\ref{app:tool_descriptions} for details on the \works present at the workshop. 
    

    \paragraph{Many Consumers, Few Executors: }
    Many of the \works presented attempt to control, or at minimum monitor,
    computational tasks executed on a physical local, HPC, or Cloud hardware,
    resulting in a barrier to mixing tools together. The central, often time
    consuming, step in BMS and broader computational molecular sciences is the
    computation itself, not necessarily moving between different tasks within
    a workflow. The execution of those computations, whether they be on local
    hardware, submitting jobs to a cluster or supercomputer, or allocating
    resources on the cloud, seem to be executed by many of the tools presented
    at the workshop. This appears to be one of the large barriers to having
    more modular tools which users can swap to suit their needs. Any tool
    which takes exclusive ownership of executing the computation typically
    cannot interface with another tool doing the same action without either
    custom implementations or intimate developer/user interaction. Although no
    solution was decided upon at the workshop, the sense was that existing
    \works should be flexible enough to allow another application to handle
    computation management and future \works should either choose to
    specialize in computation management (e.g. PyComps or Radical-CyberTools),
    or integrate a separate, preexisting tool into their workflow design.
    
\paragraph{Less is More: }    Several tools have emerged which specialize in
    heterogeneous HPC computation management. These are tools whose design
    goal is to manage job submissions on multi-user, HPC platforms typically
    managed by third-party groups. The platforms these tools run on include
    local hardware clusters and government-run supercomputers, for example,
    and often have commercial job queuing systems such as SLURM, LSF, PBS, or
    Torque. The end-user then does not have to concern themselves with writing
    submission scripts, managing jobs, collecting inputs and outputs, nor
    learning the process all over on the next platform. As HPC systems and
    cloud computing become more popular, the value of these tools can
    intuitively be expected to rise. However, other \works cannot take full
    advantage of these HPC computation management specialists if they
    themselves try to handle the management, alluding to the issue raised
    earlier.
    
    \paragraph{Python -- Pervasive and Powerful: } Python has been adopted as
    the primary language most users interface with these \works. The only
    exception to this present at the workshop was CWL, which had users provide
    a YAML file to structure their workflow. Python's prevalence is to be
    expected, given how popular it is within the computational sciences as a
    whole. The inherent problems with the language were also well understood
    in that for true HPC execution it has inefficiencies with speed, memory
    management, and parallelization. However, these problems can be addressed
    through more efficient back-ends which Python can wrap around. Overall,
    the middleware layers that the user formally interfaces with and
    ultimately writes their workflow with was most commonly Python and the
    workshop attendees foresee no changes in this trend for the near future.    
    
    \paragraph{Generalized versus Specialized: }
    Classifying the different \works presented into two fundamental categories was an issue of debate and no consensus was reached, 
    but the workshop attendees desired to incorporate a discussion of the matter. The first category was users who want a workflow to execute 
    an arbitrary sequence of dependent tasks. This includes tools like CWL and Parsl which are designed for general purpose applications. 
    The second classification is \works which are applied to solve a specific class of scientific problems; tools such as Adaptive MD, gmxAPI, 
    and some of the BioSimSpace works. One debatable observation is the second class is just a specialized application of the first class, 
    maybe even optimized for a specific scientific problem. In practice, the distinction may not matter or be practical 
    for the purposes of grouping different \works together, but instead may only serve to highlight that a user should ask 
    the questions outlined in Section~\ref{sec:tool_summary} and keep in mind that a more specialized tool may better suit their needs 
    than a general one.

\section{Recommendations we have for the community as a whole}\label{sec:recommendations}
	
	{\bf Researchers should first find a list of existing \works to draw from before choosing one tool.} A user who does so will have 
	a better understanding what features each \works have from the onset and likely find one which meets their needs.
	This is in contrast to the normal starting point where users make scripts first, then find a workflow which can string the 
	scripts together later. 
	We encourage the communities to first seek out existing \works as possible solutions to automate their workflow.
	Below are two non-exhaustive examples of curated lists of available tools. These were not created nor maintained by the authors and 
	only serve as examples.
	\begin{itemize}
	    \item https://s.apache.org/existing-workflow-systems (Last accessed Jan 2019)
	    \item https://github.com/pditommaso/awesome-pipeline (Last accessed Jan 2019)
	\end{itemize}
	Perusing lists like these will help find the \work that suits a user's needs.
	Readers are encouraged to review the workflow considerations presented in Section~\ref{sec:tool_summary} while looking at existing \works.
	Any questions regarding specific tools should be raised with the tool developers. Other organizations, such as MolSSI and BioExcel, are available
	to provide assistance in directing users to the correct contacts for individual \works.
	    
	
	{\bf Developers of \works should speak with HPC resource maintainers to try and make their tool more forward 
	facing on their system.} The product of these efforts can be items such as user guides, information on HPC websites, recommendations given 
	from the HPC teams to users, etc. Developers can pitch any given \work as a product which will provide reproducible execution of a users pipeline 
	and reduce the time required for both the user and the HPC administrators. From the user perspective, complex sequences of tasks 
	become a series of programmed instructions which will likely handle HPC resource management and task submission (as was observed 
	in Section~\ref{sec:discussion_breakdown}); this allows the user to focus on their scientific efforts, not hardware and software 
	management. From the HPC administrator side, the \work will likely handle the job and resource management of the HPC platform for the 
	user, which will reduce the amount of time administrators have to spend on technical support; this frees the administrators to focus 
	on actual HPC management, upgrades, and maintenance. This recommendation, however, runs the risk of favoring one tool over another, 
	which should not be the goal of any \work. Each tool will have its own strengths and weaknesses by design, and so developer's 
	individual \work should identify these and help guide the correct users to themselves, while 
	discouraging users for whom their \work is not suited. Consider the questions raised in  
	Section~\ref{sec:tool_summary} as guides; encourage users to answer them first and work with HPC teams to get those answers 
	from users as a precursor to presenting a specific \work.
	
	
	{\bf \work developers must take their users' needs into consideration when they write documentation and marketing for their tool.}
	The importance of \works will increase as workflows continue to grow in complexity.
	Unfortunately, there are still weaknesses in effective marketing of \works.
    One major issue is determining which \work should be used to execute a particular workflow.
    There are many available tools, but it is unclear what the strengths and weaknesses are for each particular tool.
    The documentation provided by a \work must provide the user with sufficient information that they may determine the applicability of the tool to their specific workflow.
    In an effort to aid that decision, this report has provided a set of considerations to keep in mind
    when a user is looking into \works, but there are also responsibilities on the tool developers
    (see Section~\ref{sec:tool_summary}).
    Each \work must provide clear documentation for their tool to aid users in applying it and to help
    users decide if the tool is right for their needs in the first place.

    {\bf \work developers need to increase communication and interaction with other tool developers.}
    \works are being developed too much in isolation of one another, leading to many teams encountering
    similar problems and handling them independently.
    Many \works have a vertical integration model, meaning a single tool is performing all the operations associated with the workflow, such as building the workflow, determining dependencies and task ordering, and executing the computation tasks on various compute resources.
    If more \works tried to use a less vertical model, the burden of development would distribute across more researchers with a wider range of expertise.
    A prime example of this is in task distribution systems.
    Most \works execute the steps of a constructed workflow, either locally or on a remote resource, and perform this through their own implementation of a task distribution system.
    A more productive method would be to seek out developers of distributed computing engines either to collaborate in the development of an engine for the \work or to directly apply their engine.

\section{End-User-Centric Workshop Plan} \label{sec:future-shop}
    
    A follow up workshop should be planned where several novice and potential end-users of \works are
    invited along side the 
    \works developers themselves. "User" in this context is either an individual person, a
    representative of a group, or 2-3 members of 
    a group who want to either adopt a \work  or see if there is a better \work for their scientific
    problem. This workshop should not 
    target users who are already versed in a given tool and have no desire to change tools. Those users
    should reach out to their 
    chosen \work developers on their own. The overall structure of the workshop will be to get user-specific pipelines operating though a paired \work with the help of the developer in an intimate, hands-on setting.
    
    This workshop will serve several larger goals for the broader workflow community beyond the immediate benefit to attendees. These goals 
    are more informative to the requirements of users which the developers are, or are not, serving and are as follows:
    \begin{itemize}
        \item Open a direct line of communication between users and developers as the users begin to apply a \work to their project. A study of this process will not only improve the attendees understanding of each sides requirements, but also provide non-attendees information about what users search for in a \work and how they conduct their search. Developers can then improve their products to better help the end users. 
        
        \item Capture the requirements of end users for workflows more generally, and the capabilities of existing \works to support them.
        This will help inform the developer community how to improve and design \works which better serve the users.
        
        \item Capture details on what advertisement and marketing techniques work for users: How are people finding these tools, how did they hear about them, did they hear about them, what would they have preferred/what would have worked? 
        
        \item Gain a better understanding of how user needs are supported, allowing \works to be brokered to users.
  
        \item Compare and contrast software gaps in the workflow community which either end-users or developers identify and attempt to document why the gaps exist.
    \end{itemize}

    
    Prerequisites for the workshop will require input from both the developers and the users as well. They will need to happen in the following 
    order to best serve the users:
    \begin{enumerate}
        \item The questions in Section~\ref{sec:tool_summary} will be explicitly requested from the \work developers invited to the workshop.
        \item The same questions will be asked of the users in the context of the specific scientific problem they are working on (or plan to).
        \item The workshop organizers, likely MolSSI and/or BioSimSpace will compare the answers from both and try to match users to specific \work.
        \item Limit the pairs so each user has a sufficient number of developers in attendance to divide and work exclusively with a user.
        \item Inform each pair who they are matched with to give them a chance to learn about one another.
        \item Have users try to implement their tool ahead of time, or at minimum have each developer provide the users with a "try to get to this point" of implementation before the workshop. 
    \end{enumerate}
    
    Once prerequisites are gathered, the workshop will be formatted in the following manner:
    \begin{itemize}
        \item Users will bring with them the specific scientific problem they are looking to apply to the \work we have paired them with, after having attempted to implement it on their own themselves. 
        \item Developers will bring some generic toy problems for users to play with. This is done because we should not expect to perfectly pair users/developers based on their answers to Section~\ref{sec:tool_summary}'s questions: users may not know the answers to some questions, there may not be enough developers to pair with each user, or we could simply get the pairing wrong.
        \item Day 1: Devote to introductions and having everyone do a hands-on work through of the developer toy problems. This will give everyone an introduction to the tools available and a feel for their differences and commonalities.
        \item Day 2: Split the group into their pairs and have each pair work to implement the specific scientific problems the users bring within the matched \work from the developer.
        \item Day 3: Flexible day where users can request to work with another \work developer in a slightly less intimate setting as they may feel another \work will suit their needs better.
        \item Wrap up: Gather feedback from the workshop. Focus on what the next major steps are. If a similar workshop were to be held at a larger-scale conference (e.g. SuperComputing), what would it look like? What should be changed? Was the workshop helpful?
    \end{itemize}
    The days are flexible here, e.g the one-on-one day could be split over two whole days, or one afternoon and one morning.
    
    There are two major risks which need to be worked out: How to reach users,
    how to avoid isolation over the continents. One idea floated was to have two
    conferences: one hosted by MolSSI in the United States and another by
    BioExcel and/or BioSimSpace in Europe. Another possibility is offsetting
    some costs by requesting a European focused organization sponsor European
    users' travel expenses. More details for expenses will be worked out in
    the planning phases for the workshop. Reaching different potential users
    will be another limiting factor. Effectively this will be a customer
    discovery and outreach process and we are open to ideas how to reach
    out and advertise such a workshop, especially since workshop size will be
    a function of how many users every group of attending \work developers can
    support.

\section*{Acknowledgement}
The authors would like to acknowledge the workshop organizers, participants, and contributors. They would also like to thank the Molecular Sciences 
Software Institute, BioExcel, and the NSF ExTASY for sponsoring and funding the event. Finally. they would like to thank the Barcelona Supercomputing 
Center for hosting and organizing the event.

\newpage

\appendix
\section{Glossary of Terms}
\begin{itemize}
    \item \textbf{Computational Campaign} -- Computational activity designed to achieve a specific goal.
    \item \textbf{Workflow} -- A specified and well-defined set of computational tasks to be executed. Often a computational campaign is made up of one or more workflows.
    \item \textbf{\work} -- A tool used to construct and deploy a workflow.
    \item \textbf{Workload} Discrete unit of a computational task that can be executed, devoid of semantic context. Sequential Workloads comprise a Workflow.
\end{itemize}

\newpage 

\section{\work Descriptions}\label{app:tool_descriptions}

    This section documents the workflow systems who were in attendance at the workshop in alphabetical order. Each sub section includes a brief
    description of the workflow provided either by the developer or from the tool's website. 

    \subsection{AdaptiveMD}
        AdaptiveMD is a Python package designed to create HPC-scale workflows (104 parallel tasks) for adaptive sampling of biomolecular MD simulations. This method seeks to improve sampling of the slowest processes, i.e. those on biologically relevant timescales, with unbiased MD replicates. To run reliably and independently over the weeks to months typical of a general use-case, AdaptiveMD was designed as a distributed application that can run from a laptop or directly on an HPC resource and automate asynchronous workflow creation and execution. Multiple adaptive sampling algorithms are fully automated with minimal user input, while advanced users can easily make modifications to workflow parameters and logic through the Python API. Users can prototype and utilize sampling algorithms via their own (potentially very small) analysis and frame selection scripts with simple a simple interface to the MD trajectory data structure. Our current out-of-the-box restart state adaption can be expanded to utilize interim data from the workflow (or arbitrary logic) for making runtime adaptions to 1) other task properties such as analysis type or parameters, 2) workload properties such as task count, or 3) workflow properties such as convergence criteria. 

        Validation cases are currently running on OLCF Titan to compare adaptive sampling workflows with very long trajectories run on the specialized Anton MD supercomputer \cite{lindorff}. AdaptiveMD is also currently being used for novel investigations on OLCF Summit and Titan to elucidate mechanisms of lignocellulosic biomass decomposition in various solvent conditions, and to characterize medically relevant differences in disease-implicated protein mutants for the MHCII antigen presentation complex and IMPDH protein. 

        AdaptiveMD can address the entire chain from a workflow-generating instance down to task execution when using its native worker class. This paradigm is sufficient for rapid deployment on small-scale or lab specific resources, where dedicated workflow management software is not available or otherwise accessible to a user. To provide robust workflow management AdaptiveMD is also integrated with the Radical Cybertools stack, which greatly enhances the runtime error detection and correction functionality, but has a much higher installation and configuration overhead if not already available on a user’s resource. To keep configuration and installation simple while providing reliable workflow execution on a wide range of computational resources, further development efforts will focus on 1) isolating the workflow-generating functionality and 2) integrating with multiple backends that can be seamlessly activated for task execution on a wide variety of resources.

    \subsection{Crossbow and Crossflow}
        Crossbow is a Python-based toolkit to provide an easy entry to cloud-based computing for biomolecular simulation scientists. 
        It is particularly aimed at end users with limited expertise in system administration. Crossbow provides “one click’ deployment 
        of compute clusters consisting of a head node and variable number of worker nodes, all linked with a shared file system. 
        Crossflow is a Python-based toolkit for workflow construction and execution, aimed particularly at Crossbow clusters but more 
        generally at distributed computing environments. 

        Crossflow shares many of its design aspects with Parsl. It provides tools to wrap Python functions and external applications 
        (e.g. legacy MD simulation codes), in such a way that they can be combined into workflows using a task-based paradigm. Crossflow 
        uses Dask Distributed as the task scheduling and execution layer. This handles the construction of the task graph, optimal 
        scheduling of tasks on resources (workers), data distribution, and resilience. Though Crossflow has so far mainly been tested 
        on Crossbow distributed clusters, it will run on any set of resources on which Dask Distributed can be deployed, which 
        includes traditional HPC type systems.

        Crossbow/Crossflow is a relatively new (2 years old as of Jan. 2019) project, but demonstrator versions of a wide range of 
        workflows relevant 
        to the biomolecular simulation community have been developed. These include workflows for system preparation (high throughput, 
        standardized workflows for e.g., system parameterization, solvation and equilibration), enhanced sampling of conformational 
        space (e.g. simulation/analysis loops) and replica exchange methods (both temperature replica exchange and more general 
        Hamiltonian replica exchange). The molecular modelling and simulation tools that have been interfaced with Crossflow include a 
        variety of standard MD codes (Amber, Gromacs, and NAMD), validation tools (Whatcheck), analysis tools (ProPka, FPocket), 
        and docking tools (AutoDock Vina).

    \subsection{Common Workflow Language (CWL)}
        \textit{Authors' Note: The following description is taken from the CWL main website \cite{cwl_site}}
        
        \begin{quote}
            The Common Workflow Language (CWL) is a specification for describing analysis workflows and tools in a way that makes them portable and
            scalable across a variety of software and hardware environments, from workstations to cluster, cloud, and high performance computing (HPC)
            environments. CWL is designed to meet the needs of data-intensive science, such as Bioinformatics, Medical Imaging, Astronomy, Physics, and
            Chemistry.
    
            CWL is developed by a multi-vendor working group consisting of organizations and individuals aiming to enable scientists to share data analysis
            workflows. The CWL project is maintained on Github and we follow the Open-Stand.org principles for collaborative open standards development.
            Legally CWL is a member project of Software Freedom Conservancy and is formally managed by the elected CWL leadership team, however every-day
            project decisions are made by the CWL community which is open for participation by anyone.
    
            CWL builds on technologies such as JSON-LD for data modeling and Docker for portable runtime environments.
        \end{quote}

    \subsection{Fireworks}
        FireWorks is a free, open-source code for defining, managing, and executing workflows. Complex workflows can be defined using Python, JSON, or YAML, are stored using MongoDB, and can be monitored through a built-in web interface. Workflow execution can be automated over arbitrary computing resources, including those that have a queueing system. FireWorks has been used to run millions of workflows encompassing tens of millions of CPU-hours across diverse application areas and in long-term production projects over the span of multiple years. In particular, it has been used by many materials science (i.e. Materials Project, JCESR, JCAP), chemistry, and catalysis research. It has been reported to be used also for graphics processing, machine learning, multi-scale modeling, and document processing. FireWorks has a very active support provided through its forum directly by some of its developers and other expert users. Particular attention is spent in writing its documentation. It results to be friendly for a new user and technical for expert users at the same time. An academic paper on FireWorks is also available.

        Some features that distinguish FireWorks are dynamic workflows, failure-detection routines, and built-in tools and execution modes for running high-throughput computations at large computing centers. While FireWorks provides many features, its basic operation is simple. There are essentially just two components of a FireWorks installation: a server (“LaunchPad”) that manages workflows and one or more workers (“FireWorkers”) that run your jobs. You can add workflows (a DAG of “FireWorks”) to the LaunchPad, query for the state of your workflows, or rerun workflows. The FireWorkers (a single laptop or at a supercomputing center) request workflows from the LaunchPad, execute them, and send back information.

        Workflows in FireWorks are made up of three main components. A Firetask is an atomic computing job. It can call a single shell script or execute a single Python function that you define. A Firework contains the JSON spec that includes all the information needed to bootstrap your job. The spec also includes any input parameters to pass to your Firetasks. A Workflow is a set of FireWorks with dependencies between them. Between FireWorks, you can return a FWAction that can store data or modify the Workflow depending on the output (e.g., pass data to the next step, cancel the remaining parts of the Workflow, or even add new FireWorks that are defined within the object).

    \subsection{gmxAPI: Gromacs API}
        The gmxAPI is a Python wrapper which links to Gromacs for all cases where people would otherwise write scripts to call Gromacs binaries and 
        functions. This includes users who are scripting Gromacs jobs for HPC, methods developers who are wrapping Gromacs calls in their methods, or methods/software developers who are currently doing custom hooks to Gromacs. The user base is intended to be those with some scripting 
        and program skills, but do not need to be advanced users of GROMACS or molecular dynamics and workflow software.
        
        gmxAPI's interface is itself not limited to job size. Developer use cases are on $\mathcal{O}(100)$ simultaneous jobs through the 
        MPI-based runner. Job start and stop latency is not the limiting factor so the individual nature of the jobs and ensemble execution 
        back-ends will be the limiter; these limits will be user and specific scientific problem dependent. The current back-end task 
        runner operates on MPI, and there is a desire to provide additional back-ends for larger scale tasks, and make it so users 
        can develop their own backends as well.
        
        The wider user community of GROMACS itself will have access to the initial versions of gmxAPI as the tool set will be integrated
        in GROMACS. This community is a self-sustaining one which will help gmxAPI grow through feedback and development.

    \subsection{Parsl}
        Parsl is a Python library for programming and executing data-oriented workflows (dataflows) in parallel. Parsl scripts allow selected Python functions and external applications (called apps) to be connected by shared input/output data objects into flexible parallel workflows. Rather than explicitly defining a dependency graph and/or modifying data structures, instead developers simply annotate Python functions and Parsl constructs a dynamic, parallel execution graph derived from the implicit linkage between apps based on shared input/output data objects. Parsl then executes apps when dependencies are met. Parsl is resource-independent, that is, the same Parsl script can be executed on a laptop, cluster, cloud, or supercomputer.
    
        
        Parsl is used for a variety of data-oriented workflows ranging from traditional many task computing workflows through to newer online, machine learning, and interactive computing models. In each case the underlying workflows share similar structures that include pipeline (series of connected applications) and more complex conditional/loop-based workflows (e.g., simulation-analysis loop). These workflows arise in many scientific domains including biology, physics, cosmology, chemistry, and social sciences.  

        Parsl and its predecessors, Swift/K and Swift/T, have enabled a wide variety of scientific successes including simulation of super-cooled glass materials, protein and biomolecule structure and interaction, climate model analysis and decision making for global food production and supply, materials science at the Advanced Photon Source, multiscale subsurface flow modeling, power grid modeling, high resolutions surface modeling of the arctic, large scale neural network hyperparameter optimization for cancer research, understanding the physics of overtaking maneuver in Indy Car racing, high resolution modeling of urban airflow. using machine learning to predict stopping power in materials, study of ionic liquids and deep eutectic solvents  information extraction, and in the simulation of cosmic ray showers for high school students.
        
        Recently Parsl has been used to simulate images to be obtained from the Large Synoptic Survey Telescope (LSST). This simulation is crucial for developing the workflows needed to analyze LSST data with the aim to measure how Dark Energy behaves over time. Members of the Dark Energy Science Collaboration have created a Parsl-based workflow that executes a variety of image reconstruction steps based on simulated instance catalogs. The workflow uses Singularity containers that hold the primarily Python-based simulation code. Parsl then orchestrates execution of a series of commands by first bundling tasks into appropriate sizes for nodes (considering available resources). It then runs over 10,000s of instance catalogs, each with 189 simulated sensors to create images. The workflow has been run on Argonne's Theta supercomputer and NERSC's Cori supercomputer. In one run, the workflow used 4,000 Theta nodes (256K cores) for 72 hours. 

    \subsection{PyCOMPSs}
        
        PyCOMPSs/COMPSs is a task-based programming model for the development of workflows/applications to be executed in distributed programming platforms. 
        The workflow is described with a sequential Python code (or Java or C/C++) with tasks annotated in the form of Python decorators. An important aspect is to indicate the directionality of the parameters (input, output or inout). At execution time the actual data-dependencies between tasks are derived by the runtime, that is why the workflow is actually dynamic. The syntax includes a tiny set of synchronization calls, which its use should be minimized to reduced global synchronization points. Internally tasks can be sequential or parallel (intra- and inter-node parallelism is supported). The runtime is written in Java for historical reasons. Has several components with interfaces more or less well defined. Components such as the scheduler can be configured at execution time to choose a locality aware scheduler, a fast scheduler, etc. The runtime performs the graph creation, task scheduling, resource management and data management. Interoperability with different type of computing resources is provided by the runtime as well (besides HPC systems, clouds and federated clouds are supported and clusters and clouds managed by container engines). 
        
        
        PyCOMPSs is based on general purpose programming languages and focuses on being an enabler for multiple science areas: life science, earth science, astrophysics, etc. Our focus is to abstract infrastructure to the workflow (make application agnostic of the actual computing platform). In this sense, the code does not include explicit parallelization directives, neither calls to APIs for accessing the platform resources, etc. The COMPSs runtime is able to automate the parallelization of the codes at task based approach, concurrency detected at execution time, and performs other features, like data transfer and resource management in an autonomous way. The runtime makes scheduling decisions and chooses where/when to execute each workflow node. The   ame code can run on clusters, grids, clouds, clusters managed with container managers. PyCOMPss offers an abstraction of the storage distribution or memory address space to the applications. 
        
        PyCOMPSs focuses on offering an environment that supports dynamic workflows. The actual workflow is instantiated automatically at execution time, and its morphology depends on input data. 
        PyCOMPSs supports different flavours:  HPC orientation: combination of automatic parallelization + traditional parallel codes; Service orientation: each node task can be a previously deployed web service; also the whole application can be a web service;  Support for mobile/edge devices: devices can appear/disappear dynamically.
        
        
        PyCOMPSs/COMPSs has been successfully applied to multiple application areas, such as life-science, earth-science, astrophysics, biodiversity, etc. We have developed a large number of use-cases in the multiple funded projects that we have participated. We are especially proud of two recent developments in the area of life sciences. The Guidance application, that solves the GWAS problem. We have been working with the life-science department at BSC in the development of this application, with successful scientific results in the research for the genomic origins of type-2 diabetes. The Guidance application has been run with around 10.000 cores in the MareNostrum 4 supercomputer.  Another success in this area are the development of the BioExcel Building Blocks on top of PyCOMPSs. With our colleagues at the life-science department, we have developed the PyMDsetup workflow which has been run with around 10.000 cores and a new run with around 40.000 cores is planed this month. 
        Other successful applications are in the area of Artificial Intelligence (integrating with TensorFlow) or climate (supporting workflows of MPI simulations). 
        
        
        
        

    \subsection{RADICAL-Cybertools (RCT)}
        RADICAL-Cybertools enable the execution of ensemble-based applications on a variety of high performance computing infrastructures. An increasing number of scientific domains is adopting and benefiting from ensemble-based applications. Most notably, Molecular Dynamics is moving away from implementing molecular simulations as a single, long-running, very large MPI job to a large set of shorter-running, small to medium size simulations executed concurrently. This shift is explained by both scientific and algorithmic insight. Scientifically, a statistical approach to exploring the phase space of molecular behavior is proving to be more effective at avoiding typical pitfalls like missing rare but meaningful events or stalling in a specific region of the phase space. Algorithmically, ensemble-based applications can benefit from weak scaling where traditional single simulations need strong scaling, made increasingly more difficult from the capping of single-core performance.
        
        RADICAL-Cybertools (RCT) consist of four independent software systems: RADICAL-SAGA (RS), RADICAL-Pilot (RP), RADICAL-EnsembleToolkit (EnTK), and RADICAL-Analytics (RA). RS enables interoperability across job schedulers, file transfer and resource provisioning services via a unified, high-level API. RP implements the pilot paradigm and architectural pattern offering concurrent execution of heterogeneous tasks on the same pilot and the support of more than twelve methods to launch tasks. EnTK allows users to develop ensemble-based applications in terms of static or adaptive pipelines, stages and tasks, abstracting away the complexities of resource and distributed execution management. Finally, RA offers a low-level API to profile traces of other RCT tools, enabling time-series analysis of each component behavior. RCT successfully enabled bio*, earth and climate science simulations from more than twenty research groups on leadership-class and large HPC machines in United States, Europe and Japan.

\end{document}